\documentclass[reprint,amsmath,amssymb,aps,prb,showkeys,nofootinbib]{revtex4-2}
\usepackage[utf8]{inputenc}
\usepackage{hyperref}
\usepackage{listings}
\usepackage{graphicx}
\usepackage{float}
\usepackage{subfigure}
\usepackage{color}

\begin{document}

\title{Symmetry breaking at a topological phase transition}
\author{Michael F. Faulkner}
\email{michael.faulkner@warwick.ac.uk}
\affiliation{HH Wills Physics Laboratory, University of Bristol, UK}
\affiliation{Warwick Centre for Predictive Modelling, University of Warwick, UK}
\date{\today}

\begin{abstract}
Spontaneous symmetry breaking is a foundational concept in physics.  In condensed matter, it characterizes conventional continuous phase transitions but is absent at topological phase transitions such as the Berezinskii--Kosterlitz--Thouless (BKT) transition -- as in the BKT case the expected norm (i.e., the magnitude) of the $U(1)$ order parameter vanishes in the thermodynamic limit at all nonzero temperatures.  Phenomena consistent with low-temperature broken symmetry have been observed, however, in many different BKT experiments.  Examples include recent experiments on superconducting films and the seminal work on two-dimensional arrays of Josephson junctions.  While the inaccessibility of the above thermodynamic limit partially explains this paradox in finite systems, the full dynamical framework of symmetry breaking at the BKT transition remains unresolved.  Here we provide this by introducing the broader concept of \emph{general symmetry breaking}.  This encompasses both spontaneous symmetry breaking and the BKT case by allowing the expected norm of the order parameter to go to zero in the thermodynamic limit, provided its directional phase fluctuations are asymptotically smaller.  We demonstrate this asymptotically slow directional mixing in the low-temperature BKT phase.  This explicitly shows that the order parameter arbitrarily chooses some well-defined direction in the thermodynamic limit, predicting negligible phase fluctuations compared to the expected norm in arbitrarily large experimental BKT systems.  Our results provide a model for directional mixing timescales across the diverse array of experimental BKT systems.  We suggest various experiments.

\begin{center}
{\it This paper is dedicated to the memory of Mr (Rob) Oldcorn -- \\ the greatest chemistry teacher and most fundamental source of scientific inspiration.}
\end{center}
\end{abstract}



\maketitle

Topology and symmetry are two of the most fundamental concepts in physics.  The former classifies objects via properties that are preserved under continuous deformation.  It provides a framework for a multitude of physical phenomena, from topological insulators~\cite{Hasan2010ColloquiumTopologicalInsulators} and the quantum Hall effect~\cite{Laughlin1981QuantizedHall} to fault-tolerant quantum computation~\cite{Kitaev2003FaultTolerant} and knots in light fields~\cite{Dennis2010Isolated} and liquid crystals~\cite{Machon2013Knots}.  It was also fundamental to the defect-mediated description of topological phase transitions in condensed matter~\cite{Berezinskii1973DestructionLongRangeOrder,Kosterlitz1973OrderingMetastability,Haldane1983NonlinearFieldTheory}, where the concept of symmetry --  the preservation of system properties under symmetry transformations -- has been possibly most powerful in the characterization of conventional 
continuous phase transitions.  For example, at low temperature, the two-dimensional (2D) Ising model restricted to single spin-flip dynamics breaks its global $Z_2$ symmetry by arbitrarily choosing either a positive or negative magnetization -- the global $Z_2$ order parameter -- and keeping this sign on a timescale that diverges with system size~\cite{Onsager1949Discussion}.  However, it is the simplest case of continuous symmetry -- the $U(1)$ group of planar rotations -- that showcases some of the most interesting and subtle symmetry properties, but at a \emph{topological} phase transition.  

In an ergodic $U(1)$ system, the directional phase $\phi_{\bf m} \in (-\pi, \pi]$ of the global $U(1)$ order parameter ${\bf m} = (\|{\bf m}\|, \phi_{\bf m})$ ergodically explores $(-\pi, \pi]$ on some finite \emph{directional mixing timescale}: the mean phase converges to its expected value $\mathbb{E} \phi_{\bf m} = 0$ on this timescale, where it is independent of global rotations due to its fluctuations also converging to their expected value $\sqrt{\mathbb{E} \phi_{\bf m}^2} > 0$ [the expected value $\mathbb{E}f := \int f({\bf x}) \pi({\bf x}) d{\bf x}$ of some observable $f({\bf x})$ is that predicted by the Boltzmann distribution $\pi({\bf x}) \propto e^{-\beta U({\bf x})}$, with $\beta$ the inverse temperature and $U$ the potential].  If symmetry is broken under the chosen dynamics, however, the directional mixing timescale diverges with system size.  This asymptotically slow directional mixing reflects the order parameter arbitrarily choosing some well-defined direction (with zero phase fluctuations) in the thermodynamic limit.  In most cases, this can be expressed mathematically by calculating the expected order parameter $\mathbb{E}{\bf m}$ under the influence of a fixed-direction symmetry-breaking field, before taking the thermodynamic and then zero-field limits.  The resultant nonzero vector then aligns with the direction of the field, while the thermodynamic limit is singular~\cite{Berry1995Asymptotics} because exchanging the order of the two limits returns zero.  The singular limit reflects measurable zero-field discrepancies between experimental observations and predictions of the Boltzmann model (i.e., small experimental phase fluctuations compared to their expected value).  This elegantly characterizes the dynamical zero-field phenomenon of spontaneous symmetry breaking~\cite{Beekman2019SpontaneousSymmetryBreaking} in terms of thermodynamic expectations, generalizing to any symmetry group and occurring at all conventional continuous phase transitions.  Topological phase transitions are, however, an exception, described instead in terms of a topological ordering that typically breaks ergodicity more generally~\cite{Palmer1982}.  For example, spontaneous symmetry breaking is absent at the Berezinskii--Kosterlitz--Thouless (BKT) transition~\cite{Salzberg1963EquationOfStateTwoDimensional,Berezinskii1973DestructionLongRangeOrder,Kosterlitz1973OrderingMetastability} because (in zero field) the expected norm $\mathbb{E} \|{\bf m}\|$ of the $U(1)$ order parameter goes to zero in the thermodynamic limit at all nonzero temperatures~\cite{Mermin1966AbsenceFerromagnetism,Hohenberg1967ExistenceOfLong-RangeOrder}, but a suppression (under Brownian spin dynamics) of global topological defects breaks ergodicity at this paradigmatic topological phase transition~\cite{Faulkner2015TSFandErgodicityBreaking}.

Phenomena consistent with low-temperature broken symmetry have been measured, however, on experimental timescales in a diverse array of BKT systems~\cite{Baity2016EffectiveTwoDimensionalThickness,Shi2016EvidenceCorrelatedDynamics,Bishop1978StudySuperfluid,Bramwell2015PhaseOrder,Wolf1981TwoDimensionalPhaseTransition,Resnick1981KosterlitzThouless,Bramwell1993Magnetization,Huang1994MagnetismFewMonolayersLimit,Elmers1996CriticalPhenomenaTwoDimensionalMagnet,BedoyaPinto2021Intrinsic2DXYFerromagnetism,Hadzibabic2006KosterlitzThouless,Fletcher2015ConnectingBKTAndBEC,Christodoulou2021Observation}, including on very long timescales~\cite{Shi2016EvidenceCorrelatedDynamics} and in systems that approach 
idealized $U(1)$ symmetry~\cite{Bishop1978StudySuperfluid,Bramwell2015PhaseOrder}.  This suggests that the rather elegant mathematical formalism of spontaneous symmetry breaking may be too restrictive to account for all physical observations.  For example, recent measurements of the electrical resistances of films of lanthanum strontium copper oxide (LSCO) exhibited strongly nonergodic probability density functions (PDFs) 
between the BKT and mean-field superconducting transition temperatures, in contrast with Gaussian Aslamazov--Larkin-type~\cite{Aslamazov1968InfluenceOfFluctuationPairing,Aslamazov1968EffectOfFluctuations,Larkin2005FluctuationsInSuperconductors} fluctuations involving both the amplitude and phase of the condensate wavefunction above the mean-field transition~\cite{Shi2016EvidenceCorrelatedDynamics} (the BKT transition was significantly lower than the mean-field transition, resulting in a broad temperature range dominated by BKT phase fluctuations, i.e., where amplitude fluctuations are negligible~\cite{Baity2016EffectiveTwoDimensionalThickness}).  As large condensate-phase differences (over long distances) induce resistance and the experimental timescale was on the order of ten hours, this is likely to be due to increasingly large regions of symmetry-broken (and therefore persistent on some significant timescale) condensate-phase coherence during the approach to the transition -- before an onset of distinct behavior in which a single symmetry-broken region spans the entire zero-resistance system at low temperature.  (Here, condensate-phase coherence describes positional coherence of the local phase of the condensate wavefunction.)  Similarly, zero-resistance measurements on 2D arrays of Josephson junctions also provided direct experimental evidence of system-spanning condensate-phase coherence at low temperature~\cite{Wolf1981TwoDimensionalPhaseTransition,Resnick1981KosterlitzThouless}.  Moreover, results consistent with the phenomenon have been measured in superfluid~\cite{Bishop1978StudySuperfluid,Bramwell2015PhaseOrder}, magnetic~\cite{Bramwell1993Magnetization,Huang1994MagnetismFewMonolayersLimit,Elmers1996CriticalPhenomenaTwoDimensionalMagnet,BedoyaPinto2021Intrinsic2DXYFerromagnetism} and cold-atom~\cite{Hadzibabic2006KosterlitzThouless,Fletcher2015ConnectingBKTAndBEC,Christodoulou2021Observation} films, including very recent experiments on cold atoms~\cite{Christodoulou2021Observation} and a monolayer magnet~\cite{BedoyaPinto2021Intrinsic2DXYFerromagnetism}.  Elsewhere, experiments on cylindrical arrays of superconducting qubits measured a nonzero order-parameter norm~\cite{King2018ObservationOfTopological}, consistent with broken symmetry on some timescale.  Indeed, we are not aware of a BKT experiment that \emph{contradicts} broken symmetry at low temperature. 

A partial explanation for the paradox in finite systems was provided by the expected low-temperature norm going to zero very slowly and at the same rate as its fluctuations~\cite{Archambault1997MagneticFluctuations,Bramwell1998Universality}.  This led to much success in describing magnetic-film experiments~\cite{Bramwell1993Magnetization,Huang1994MagnetismFewMonolayersLimit,Elmers1996CriticalPhenomenaTwoDimensionalMagnet,Chung1999EssentialFiniteSizeEffect,BedoyaPinto2021Intrinsic2DXYFerromagnetism,Venus2022RenormalizationGroup}, but the thermodynamic limit was not addressed, and a rigorous dynamical framework for the $U(1)$ order parameter arbitrarily choosing some well-defined direction in the thermodynamic limit (i.e., for broken symmetry) remains unresolved.  The latter is particularly pertinent to the superconducting film and Josephson-junction array, as the electrical resistance is a directly measurable quantity (on very long timescales~\cite{Shi2016EvidenceCorrelatedDynamics}) that is conjugate to the directional condensate phases.  Moreover, measurements of the magnetization vector in BKT magnetic films should provide direct experimental evidence of system-spanning symmetry-broken spin-phase coherence.  In contrast, condensate-phase coherence cannot be directly measured in the superfluid helium film as there is no conjugate field.  This is despite the theory establishing its BKT transition~\cite{Bishop1978StudySuperfluid,Ambegaokar1978DissipationInTwo-dimensionalSuperfluids} implying a significant expected low-temperature norm in macroscopic systems~\cite{Bramwell2015PhaseOrder}, consistent with the accompanying system-spanning condensate-phase coherence required for macroscopic superflow.  

Here we show that topological nonergodicity in the low-temperature BKT phase~\cite{Faulkner2015TSFandErgodicityBreaking} induces asymptotically slow directional mixing of the $U(1)$ order parameter, reflecting its phase fluctuations going to zero in the thermodynamic limit (for a model system with Brownian spin dynamics).  Moreover, the phase fluctuations are asymptotically smaller than the expected norm.  This explicitly demonstrates that the $U(1)$ order parameter arbitrarily chooses some well-defined direction in the thermodynamic limit, providing a theoretical framework for negligible $U(1)$ phase fluctuations compared to the strictly decreasing expected norm in arbitrarily large experimental BKT systems, and thus for broken symmetry to the thermodynamic limit throughout the low-temperature BKT phase.  This is an example of \emph{general symmetry breaking}, which broadens the definition to encompass all phenomena 
that result in directional (phase) fluctuations going to zero in the thermodynamic limit while being asymptotically smaller than the expected norm (these are the only requirements for broken symmetry, and are fulfilled by cases of spontaneous symmetry breaking as the expected norm is nonzero in the thermodynamic limit).  This framework is also described by a singular thermodynamic limit because taking the long-time limit of the phase fluctuations before the thermodynamic limit returns their expected value, while inverting the order of the limits returns zero.  These results demonstrate an interplay between topology and symmetry, parting with orthodox theory in the form of a topological phase transition that does in fact break symmetry, but outside the elegant yet restrictive definition of spontaneous symmetry breaking (we stress that symmetry breaking is typically absent at topological phase transitions, but that these results provide the community with a single counterexample).  This provides a model for directional mixing timescales across  the wide and diverse array of experimental BKT systems.  We also introduce the concept of long-time directional stability and infer from earlier work~\cite{Tobochnik1979MonteCarlo,Archambault1997MagneticFluctuations,Bramwell1998Universality} that the expected norm does not reach its thermodynamic value of zero at arbitrarily large system size.  A very recent renormalization-group analysis of $U(1)$ phase fluctuations in BKT magnetic films~\cite{Venus2022RenormalizationGroup} further enhances the timeliness of this article.  

The paper is organized as follows.  In Section~\ref{sec:SingularLimit}, we introduce the $U(1)$ symmetry-breaking order parameter and its phase fluctuations.  We then demonstrate that the choice of system dynamics can result in asymmetric low-temperature simulations on significant timescales, consistent with the existence of a singular thermodynamic limit of the phase fluctuations.  In Section~\ref{sec:GlobalTwistDynamics}, we define global-twist dynamics, allowing us to demonstrate that topological ergodicity ensures $U(1)$ symmetry.  In Section~\ref{sec:SymmetryBreaking}, we demonstrate that Brownian spin dynamics break $U(1)$ symmetry in the low-temperature BKT phase, but that an alternative choice of local spin dynamics ensures $U(1)$-symmetric simulations on non-divergent timescales at all nonzero temperatures -- in analogy with Swendsen--Wang/Wolff simulations~\cite{Swendsen1987Nonuniversal,Wolff1989Collective} of the 2D Ising model.  We discuss our results and suggest various experiments in Section~\ref{sec:Discussion}.

\section{Singular limit}
\label{sec:SingularLimit}

\begin{figure*}[t]
  \includegraphics[width=\linewidth]{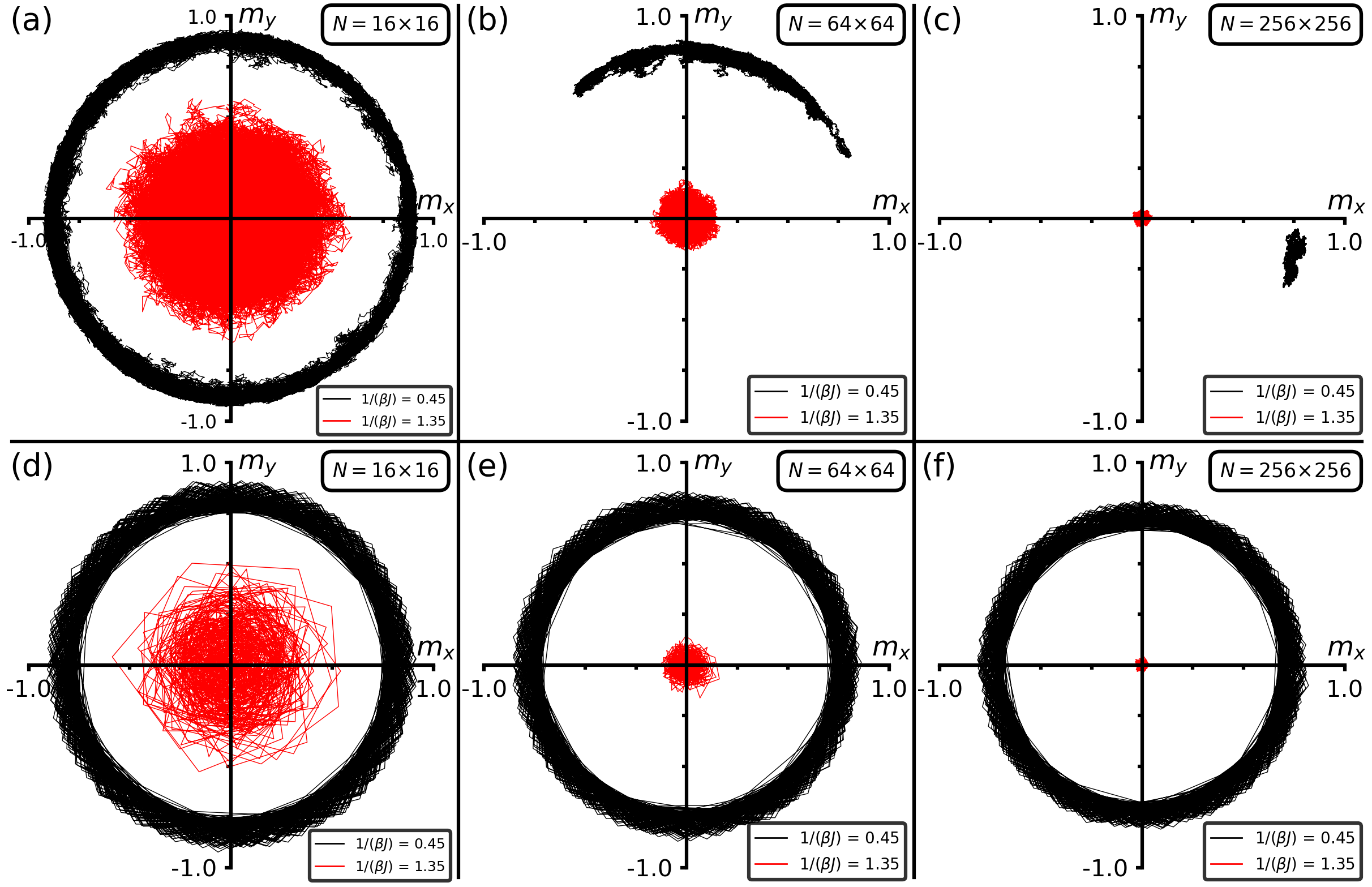}
  \caption{Evolutions of the $U(1)$ order parameter ${\bf m} := \sum_i (\cos{\varphi_i}, \sin{\varphi_i}) / N$ over the course of single Metropolis [(a)-(c)] and event-chain [(d)-(f)] simulations suggest that local Metropolis/event-chain dynamics do/do not result in the singular thermodynamic limit described below Eq.~\eqref{eq:SingularLimit}.  The local Metropolis simulations comprise $10^5$ observations with acceptance rate $a_{\rm Metrop} \simeq 0.6$.  The event-chain simulations comprise $10^3$ observations.  Straight lines connect adjacent observations.  The Metropolis data suggest a low/high-temperature directional mixing timescale that \mbox{does/does not} diverge with system size, whereas the event-chain data suggest a non-divergent directional mixing timescale for all finite $\beta$.  In particular, the low-temperature Metropolis simulations become less symmetric with increasing system size, 
  i.e., the (unbiased) simulation variance $s_{\phi_{\bf m}}^2$ of the global $U(1)$ phase $\phi_{\bf m}$ deviates further from its expected value ${\rm Var}[\phi_{\bf m}] = \pi^2 / 3$ [the variance of the uniform distribution $\mathcal{U}(-\pi, \pi)$].  This is consistent with the $U(1)$ order parameter arbitrarily choosing some well-defined direction in the thermodynamic limit.}
  \label{fig:MagnetisationEvolution}
\end{figure*}

The BKT transition governs physics as varied as 2D melting~\cite{Kosterlitz1973OrderingMetastability,Halperin1978TheoryTwoDimensionalMelting,Young1979Melting,Bernard2011TwoStepMelting,Thorneywork2017TwoDimensionalMelting}, 
2D arrays of superconducting qubits~\cite{King2018ObservationOfTopological}, Josephson junctions~\cite{Wolf1981TwoDimensionalPhaseTransition,Resnick1981KosterlitzThouless} and Bose-Einstein condensates~\cite{Trombettoni}, and planar superfluids~\cite{Bishop1978StudySuperfluid,Bramwell2015PhaseOrder},  superconductors~\cite{Baity2016EffectiveTwoDimensionalThickness,Shi2016EvidenceCorrelatedDynamics}, magnets~\cite{Bramwell1993Magnetization,Huang1994MagnetismFewMonolayersLimit,Elmers1996CriticalPhenomenaTwoDimensionalMagnet,BedoyaPinto2021Intrinsic2DXYFerromagnetism} and cold-atom systems~\cite{Hadzibabic2006KosterlitzThouless,Fletcher2015ConnectingBKTAndBEC,Christodoulou2021Observation}.  Its prototypical model is the 2DXY model of magnetism -- a set of unit-length $U(1)$ spins fixed  at the $N$ sites of a topologically toroidal, square lattice with interaction potential  
\begin{align}
U := - J \sum_{\langle i,j \rangle} \cos \left( \varphi_i - \varphi_j \right) - {\bf h} \cdot \sum_i \begin{pmatrix} \cos{\varphi_i} \\ \sin{\varphi_i} \end{pmatrix} .
\label{eq:2DXYModelDefinition}
\end{align}
Here, $J > 0$ is the exchange constant, ${\bf h} \in \mathbb{R}^2$ is the symmetry-breaking field, $\varphi_i \in (-\pi, \pi]$ is the spin phase at site $i \in \{ (1, 1), (2, 1), \dots , (\sqrt{N}, \sqrt{N}) \}$ and the sum $\sum_{\langle i,j \rangle}$ is over all nearest-neighbor spin pairs.  In zero field, the low-temperature phase is characterized by algebraic spin--spin correlations~\cite{Berezinskii1973DestructionLongRangeOrder,Kosterlitz1973OrderingMetastability} while the transition to the high-temperature disordered phase is induced by the thermal dissociation of bound pairs of local topological defects in the spin-difference field~\cite{Kosterlitz1973OrderingMetastability}, which map to charge-neutral pairs of particles in the 2D electrolyte (see Appendix~\ref{app:TopologicalErgodicity})~\cite{Kosterlitz1973OrderingMetastability,Jose1977Renormalization,Vallat1994CoulombGas,Faulkner2015TSFandErgodicityBreaking,Faulkner2017AnElectricFieldRepresentation}.  The magnetization ${\bf m} := \sum_i (\cos{\varphi_i}, \sin{\varphi_i}) / N$ is then the $U(1)$ symmetry-breaking order parameter because its expectation is proportional to the gradient of the free energy with respect to the symmetry-breaking field ${\bf h}$.  Writing ${\bf m} = (\| {\bf m} \|, \phi_{\bf m})$ in polar coordinates, we therefore define the \emph{phase fluctuations} $\langle s_{\phi_{\bf m}}^2 (\beta, \tau, N) \rangle^{1/2}$ of some 
simulation method (e.g., Metropolis) to be the root mean of the (unbiased) variances $s_{\phi_{\bf m}}^2 (\beta, \tau, N)$ (of the global $U(1)$ phase $\phi_{\bf m} \in (-\pi, \pi]$) of a large number of $N$-spin simulations of timescale $\tau$, with their expected value $\sqrt{{\rm Var}[\phi_{\bf m}]}$ given by the standard deviation ($\pi / \sqrt{3}$) of the uniform distribution $\mathcal{U}(-\pi, \pi)$.  Here, $\langle \cdot \rangle$ denotes the mean over an infinite number of independent simulations at fixed $\beta, \tau, N$~\footnote{While the phase fluctuations are defined as a statistical mean over many realizations, it can also be helpful to consider the analogous fluctuations of a single simulation or experiment.}, the simulation timescale $\tau > 0$ is the elapsed Monte Carlo time over the course of some entire simulation, and ${\rm Var}[f] := \mathbb{E} (f - \mathbb{E} f)^2$.

The Mermin--Wagner--Hohenberg theorem~\cite{Mermin1966AbsenceFerromagnetism,Hohenberg1967ExistenceOfLong-RangeOrder} predicts that the spontaneous order parameter $\lim_{\| {\bf h} \| \to 0} \lim_{N \to \infty} \mathbb{E}{\bf m}$ is zero at nonzero temperature, where it similarly predicts that the expected norm in zero field  $\mathbb{E} \| {\bf m}({\bf h} = 0) \|$ goes to zero in the thermodynamic limit (we now set ${\bf h} = 0$ unless otherwise stated).  Spontaneous symmetry breaking is therefore absent at the BKT transition.  The entire low-temperature phase is, however, critical, as the spin--spin (fluctuational) correlation length diverges for all finite $\beta > \beta_{\rm BKT}$, with $\beta_{\rm BKT}$ the phase transition.  The expected low-temperature norm does not therefore reach its thermodynamic value of zero at arbitrarily large system size, as the correlations are cut off on long length scales. 
More precisely, with $g({\bf x})$ the PDF of ${\bf m}$ and $\sigma_{\| {\bf m} \|}$ the standard deviation of $\| {\bf m} \|$, the PDF $\sigma_{\| {\bf m} \|} g({\bf x})$ of the fluctuation-normalized order parameter $\widetilde{{\bf m}} := {\bf m} / \sigma_{\| {\bf m} \|}$ maintains a well-defined low-temperature sombrero form in the thermodynamic limit because $\mathbb{E} \| \widetilde{{\bf m}} \|$ is system-size independent~\cite{Archambault1997MagneticFluctuations,Bramwell1998Universality}.  In general, for any given scalar observable, only if the ratio of its fluctuations with its expectation can be made arbitrarily small with increasing system size does there exist some finite system size at which the expectation can be considered to have reached the thermodynamic limit, which is not the case for $\mathbb{E} \| {\bf m} \|$ given this system-size independence.  All analysis of the global $U(1)$ phase $\phi_{\bf m}$ therefore holds to the thermodynamic limit.  This is also implied by the equivalence of the directional phases of ${\bf m}$ and ${\bf m} / \sigma_{\| {\bf m} \|}$.  We note also that the expected low-temperature norm itself goes to zero very slowly: $\mathbb{E}\| {\bf m} \| \sim N^{-1 / (8\pi \beta J)}$ as $\beta \! \to \! \infty$~\cite{Bramwell1993Magnetization} and $\mathbb{E}\| {\bf m} \| \sim N^{-1 / 16}$ at the finite-size transition~\cite{Bramwell1994Magnetization}.  The former was even shown to hold in an $N \sim 10^{10}$ superfluid film~\cite{Bramwell2015PhaseOrder}, while the latter implies that $\mathbb{E}\| {\bf m} \|$ would be $\sim 10^{-2}$ at the finite-size transition in a magnetic film ``the size of Texas''~\cite{Bramwell1994Magnetization}.

Fig.~\ref{fig:MagnetisationEvolution} shows evolutions of the order parameter ${\bf m}$ over the course of single simulations at various systems sizes and temperatures.  The data reflect a PDF with sombrero form at low temperature and the central well of the symmetric phase at high temperature ($1 / \beta_{\rm BKT} \simeq 0.887 J$~\cite{Weber1988MonteCarloDetermination}).  
In Figs~\ref{fig:MagnetisationEvolution}(a)-(c), we investigate the 2DXY model with Brownian (spin) dynamics by simulating the model with the local Metropolis algorithm, where each Metropolis iteration proposes a (local) single-spin perturbation and we draw observations after every $N$ such iterations, defining the Metropolis Monte Carlo time step as $\tau / n$ with $n$ the number of observations (Metropolis dynamics converge on Brownian dynamics~\cite{Neal2006OptimalScalingForPartially} with an $N$-independent physical time step that is proportional to the Monte Carlo time step, as outlined in Appendix~\ref{app:Algorithms}).  Figs~\ref{fig:MagnetisationEvolution}(d)-(f) benchmark the diffusive Metropolis dynamics against the event-chain Monte Carlo algorithm~\cite{Michel2015EventChain} as the latter induces rotations of the order parameter [and therefore $U(1)$ symmetry] on short timescales.  As outlined in detail in Appendix~\ref{app:Algorithms}, this is achieved by continuously advancing some \emph{active spin} at fixed (anticlockwise) velocity until a Metropolis rejection would have occurred.  This defines a \emph{particle event} and induces a new active spin.  
We draw observations every $N$ units of event-chain time and define the event-chain Monte Carlo time step as $\tau / n$.  The number of observations $n$ is fixed at $10^5$  and $10^3$ in the Metropolis and event-chain simulations.

\begin{figure*}[t]
  \hspace{-0.6em}
  \includegraphics[width=0.424\linewidth]{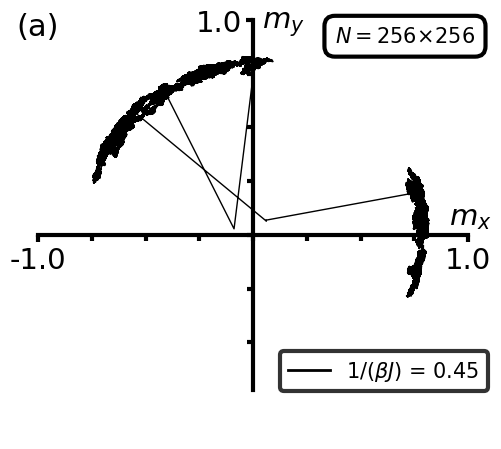}
  \includegraphics[width=0.574\linewidth]{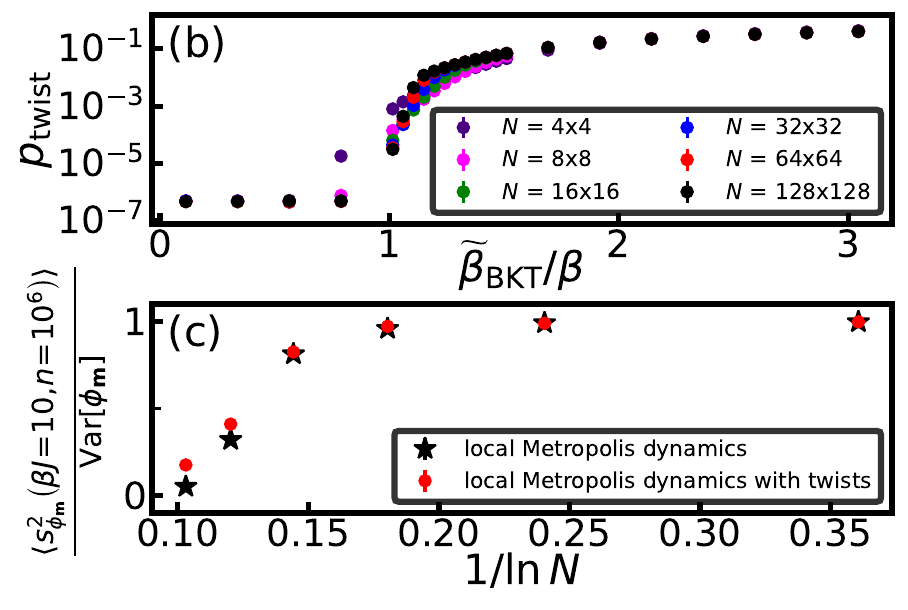}
  \vspace{-2.0em}
  \caption{Local 2DXY dynamics may be supplemented with global-twist dynamics [defined below Eq.~\eqref{eq:ExternalGlobalTwists}] that ensure $U(1)$-symmetric simulations on non-divergent timescales at all nonzero temperatures.  Global-twist events occur in pairs that form compound tunnelling events [through the sombrero potential, as in (a)] and are system-size independent sufficiently far from the transition [see (b)].  This results in zero long-time directional stability [defined in Eq.~\eqref{eq:DirectionalStability}] at all nonzero temperatures, reflected in supplemental global-twist dynamics increasing the squared phase fluctuations $\langle s_{\phi_{\bf m}}^2 \rangle$ [defined below Eq.~\eqref{eq:2DXYModelDefinition}] by an amount that increases with system size [see (c)].  Simulations use local Metropolis dynamics with acceptance rate $a_{\rm Metrop} \simeq 0.6$.  (a) Evolution of the $U(1)$ order parameter ${\bf m}$ over the course of a single Metropolis simulation (of an $N = 256 \! \times \! 256$ system) comprising $5 \! \times \! 10^5$ observations with supplemental global-twist dynamics.  (b) Probability of global-twist events vs reduced temperature $\tilde{\beta}_{\rm BKT} / \beta$ and system size $N$ based on $560 n$ attempts, with $n = 10^6$ at $\tilde{\beta}_{\rm BKT} / \beta > 1.2$ and $n = 3 \! \times \! 10^7$/$n = 10^7$ at $\tilde{\beta}_{\rm BKT} / \beta < 1.2$ for $N \gtrless 40 \! \times \! 40$ [$\tilde{\beta}_{\rm BKT} := 1 / (0.887 J)$].  (c) Low-temperature ($\beta J = 10$) squared phase fluctuations vs $1 / \ln{N}$ with and without supplemental global-twist dynamics, averaged over $5600$ simulations, each of $n = 10^6$ observations.}
  \label{fig:GlobalTwists}
\end{figure*}

In Figs~\ref{fig:MagnetisationEvolution}(a)-(c), the low-temperature Metropolis simulations are asymmetric for $N \ge 64 \! \times \! 64$, while the high-temperature Metropolis simulations appear to be symmetric for a broad range of system sizes  
(with symmetric simulations defined by a simulation variance $s_{\phi_{\bf m}}^2$ that has converged to its expected value ${\rm Var}[\phi_{\bf m}]$ within some small $\epsilon > 0$).  Moreover, the low-temperature directional mixing timescale appears to increase with system size, suggesting a singular limit.  Upon defining the \emph{long-time directional stability} 
\begin{align}
\gamma(\beta) := 1 - \lim_{\tau \to \infty} \lim_{N \to \infty} \sqrt{\frac{\langle s_{\phi_{\bf m}}^2 (\beta, \tau, N) \rangle}{{\rm Var}[\phi_{\bf m}]}} , 
\label{eq:DirectionalStability}
\end{align}
the Metropolis results are consistent with 
\begin{align}
\gamma_{\rm Metrop}(\beta) =  
\begin{cases}
    1 , \, \beta > \beta_{\rm BKT} , \\
    0 , \, \beta < \beta_{\rm BKT} ,
\end{cases}
\label{eq:SingularLimit}
\end{align}
This is due to vanishing low-temperature phase fluctuations in the thermodynamic limit, as presented in detail below.  This thermodynamic limit is singular because exchanging the order of the limits in Eq.~\eqref{eq:DirectionalStability} returns zero at all finite $\beta$ (we note that singular semiclassical limits analogously involving long times are commonplace in quantum chaos~\cite{Berry2001Chaos}).  All event-chain results in Figs~\ref{fig:MagnetisationEvolution}(d)-(f) suggest, by contrast, non-divergent directional mixing timescales, consistent with zero long-time directional stability for all finite $\beta$.  We assume local Metropolis/Brownian spin dynamics below unless otherwise stated. 

We additionally note that the low-temperature Metropolis outputs in Figs~\ref{fig:MagnetisationEvolution}(a)-(b) contain small fluctuations towards ${\bf m} = 0$ at random values of the directional phase $\phi_{\bf m}$, while the event-chain outputs in Figs~\ref{fig:MagnetisationEvolution}(d)-(f) appear to exhibit well-converged simulation variances in $\| {\bf m} \|$.  This may be a result of the asymmetry of the low-temperature $\| {\bf m} \|$ distributions in Fig. 2 of Ref.~\cite{Archambault1997MagneticFluctuations}, with the Metropolis dynamics mixing poorly in the heavy-tailed regions towards ${\bf m} = 0$.

\section{Global-twist dynamics}
\label{sec:GlobalTwistDynamics}

\begin{figure*}[t]
  \includegraphics[width=\linewidth]{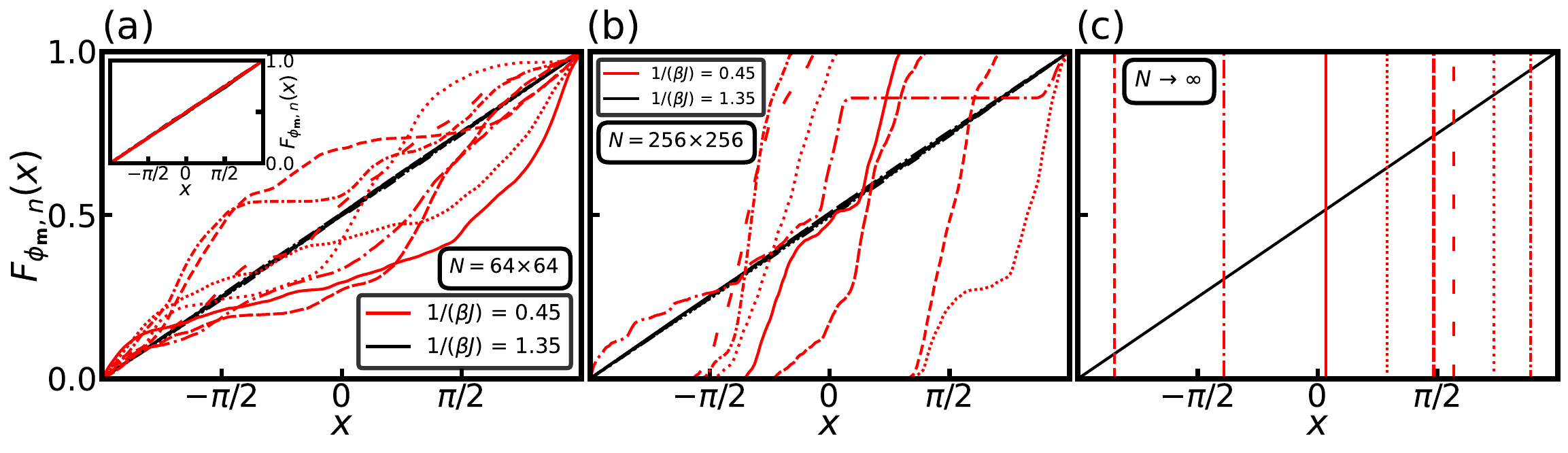}
  \caption{ECDFs [defined in Eq.~\eqref{eq:ECDF}] of the global $U(1)$ phase $\phi_{\bf m}$ under local Metropolis dynamics [(a)-(b)] reflect the strikingly different symmetry properties of the low- and high-temperature phases [red and black schematics in (c), respectively].  Different line styles represent different realizations.  (a)-(b) Eight Metropolis simulations at low and high temperature, each of $n = 10^6$ observations with acceptance rates $a \simeq 0.6$.  Results suggest that any sequence of low-temperature ECDFs at any fixed simulation timescale tends to some Heaviside step function as system size $N \to \infty$.  The high-temperature simulations suggest, by contrast, symmetric convergence on the target CDF $F(x)$ of the uniform distribution $\mathcal{U}(-\pi, \pi)$ on some non-divergent directional mixing timescale.  Inset of (a): Four event-chain simulations at low and high temperature, each of $n = 10^4$ observations.  Results are consistent with symmetric convergence on the target CDF $F(x)$ on non-divergent directional mixing timescales at all nonzero temperatures.}
  \label{fig:ECDFs}
\end{figure*}

We show below that Eq.~\eqref{eq:SingularLimit} holds for local Metropolis simulations, but we first demonstrate that zero long-time directional stability is guaranteed at all finite $\beta$ by supplementing these local dynamics with externally applied global spin twists 
\begin{align}
    \varphi_k \mapsto \varphi_k + \frac{2\pi}{\sqrt{N}} q_{x / y} k_{x/y}
    \label{eq:ExternalGlobalTwists}
\end{align}
for all spins $k \in \{ (1, 1), (2, 1), \dots , (\sqrt{N}, \sqrt{N}) \}$ along the $x/y$ dimension (${\bf q} \in \mathbb{Z}^2$).  Global-twist dynamics are then defined by one such Metropolis proposal with $q_{\mu} = \pm 1$ along each Cartesian dimension $\mu$ at each Monte Carlo time step.  Global-twist events occur in pairs that form the compound tunnelling events seen in Fig.~\ref{fig:GlobalTwists}(a), each due to a global-twist event taking the system to small $\| {\bf m} \|$ before another global-twist event returns the system to the well of the sombrero potential.  Fig.~\ref{fig:GlobalTwists}(b) shows estimates of the probability of 2DXY global-twist events as a function of temperature and system size.  The data demonstrate that, as for the case of the topological-sector events that ensure topologically ergodic simulations (defined alongside \emph{topological ergodicity} in Appendix~\ref{app:TopologicalErgodicity}) of the 2D electrolyte~\cite{Faulkner2015TSFandErgodicityBreaking}, this probability is system-size independent sufficiently far from the transition.  Moreover, the probability is non-negligible at all system sizes and nonzero temperatures (despite being small at low temperature) reflecting it scaling like $\exp (-2 \pi^2 \beta J)$ as $N \to \infty$ in the absence of other excitations.  Assuming Eq.~\eqref{eq:SingularLimit}, it then follows that, for low-temperature Metropolis simulations supplemented with global-twist dynamics, any sequence of histograms of the global $U(1)$ phase $\phi_{\bf m}$ at any fixed simulation timescale tends to some normalized sum over randomly distributed (around the well of $\widetilde{{\bf m}}$ sombrero potential) Dirac distributions in the thermodynamic limit, where a large number of Dirac distributions becomes more likely with increased (fixed) simulation timescale and we have assumed suitably chosen histogram bin sizes.  Simulations supplemented with global-twist dynamics are therefore symmetric on non-divergent timescales, with zero long-time directional stability at all finite $\beta$.  This is reflected in Fig.~\ref{fig:GlobalTwists}(c), which shows estimates of the squared phase fluctuations as a function of system size (at $\beta J = 10$) for fixed-timescale Metropolis simulations both with and without supplemental global-twist dynamics.  The data are consistent with the phase fluctuations going to zero in the thermodynamic limit for purely local Metropolis simulations, while supplemental global-twist dynamics increase the phase fluctuations by an amount that increases with system size.  As outlined in Appendix~\ref{app:TopologicalErgodicity}, supplementing local Metropolis simulations with the global-twist dynamics also ensures (and is required for) topologically ergodic low-temperature simulations on non-divergent timescales of the same order (in analogy with topological-sector dynamics in the 2D electrolyte)~\cite{Faulkner2015TSFandErgodicityBreaking,Faulkner2017AnElectricFieldRepresentation}.  Topological ergodicity therefore ensures $U(1)$ symmetry.

\section{Symmetry breaking}
\label{sec:SymmetryBreaking}

The low-temperature Metropolis data in Figs~\ref{fig:MagnetisationEvolution}(a)-(c) are consistent with vanishing phase fluctuations (at any fixed simulation timescale and under local dynamics) in the thermodynamic limit, as described by the hypothesized long-time directional stability in Eq.~\eqref{eq:SingularLimit}.  Conversely, at any fixed system size and under local dynamics, the squared phase fluctuations $\langle s_{\phi_{\bf m}}^2 \rangle$ will eventually converge to their expected value of ${\rm Var}[\phi_{\bf m}] = \pi^2 / 3$ [see Fig.~\ref{fig:GlobalTwists}(c) at small $N$] on the directional mixing timescale $\tau_{\rm mix} > 0$.  Assuming this is described by $\langle s_{\phi_{\bf m}}^2 \rangle \propto \tau / \tau_{\rm mix}$ for large enough $\tau < \tau_{\rm mix}$ under the diffusive Metropolis dynamics, the directional mixing timescale then quantifies the scaling of the fixed-timescale phase fluctuations with system size.

\begin{figure*}[t]
  \includegraphics[width=\linewidth]{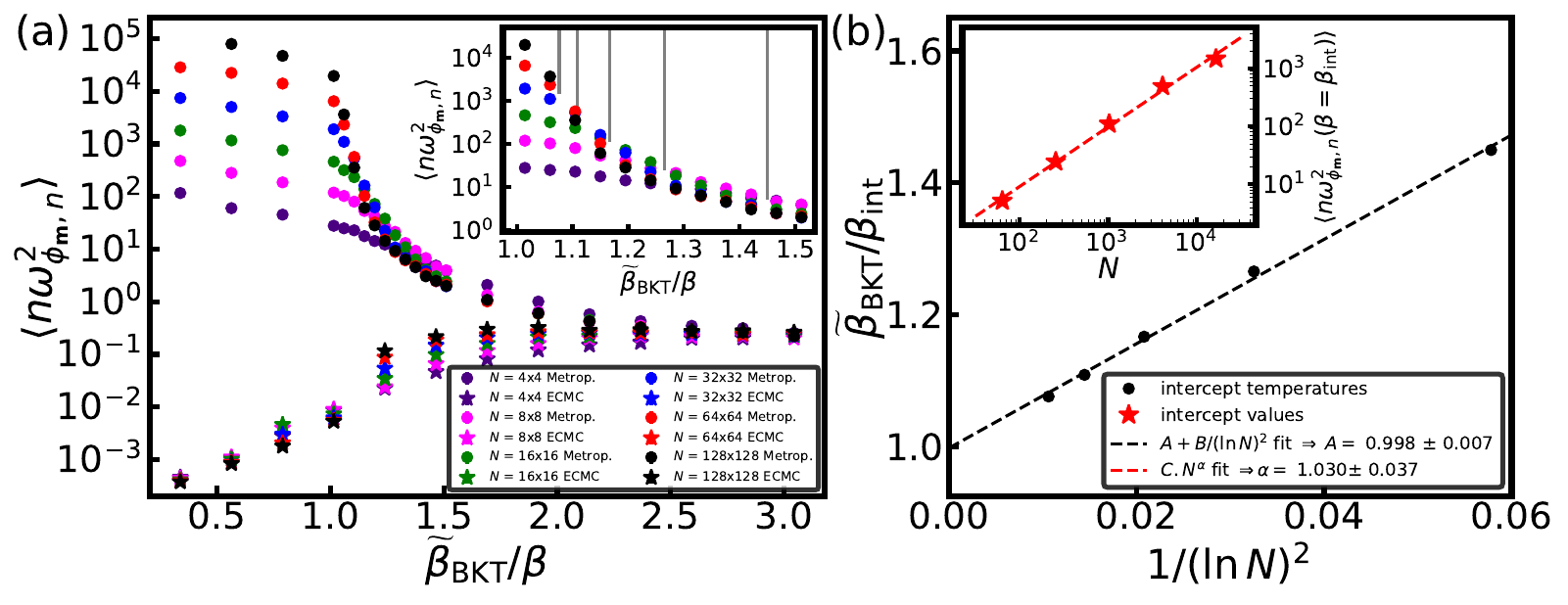}
  \caption{Brownian spin dynamics break symmetry throughout the low-temperature phase, in contrast with event-chain dynamics.  (a) Cram\'{e}r-von Mises statistic [defined below Eq.~\eqref{eq:CvM}] vs reduced temperature $\tilde{\beta}_{\rm BKT} / \beta$ and system size $N$ for Metropolis (circles) and event-chain (stars) simulations with supplemental global-twist dynamics, with $\tilde{\beta}_{\rm BKT} := 1 / (0.887 J)$.  Results indicate a directional mixing timescale $\tau_{\rm mix} \sim N^{z / 2}$ with $z = 2$ and $z = 0$ for local Metropolis and event-chain dynamics at low temperature (and $z = 0$ at high temperature in both cases).  Metropolis data sets intersect near the transition, marked by vertical grey lines in the inset.  Data is averaged over $560$ realizations with $n = 10^6$ at $\tilde{\beta}_{\rm BKT} / \beta > 1.2$ and $n = 3 \! \times \! 10^7$/$n = 10^7$ at $\tilde{\beta}_{\rm BKT} / \beta < 1.2$ for $N \gtrless 40 \! \times \! 40$.  Local Metropolis acceptance rates $a \simeq 0.6$.  (b) Estimated reduced intercept temperatures vs $1 / \left( \ln{N} \right)^2$ with a straight-line fit (black dashed line) that extrapolates to one (i.e., the phase transition) as $N \to \infty$, within the estimated error.  Inset: Estimated intercept values vs $N$ with a power-law fit indicating approximate $\sim N$ scaling.}
  \label{fig:SymmetryBreaking}
\end{figure*}

To characterize the directional mixing timescale, we plot the empirical cumulative distribution functions (ECDFs) 
\begin{align}
    F_{\phi_{\bf m},n}(x) := \frac{1}{n} \sum_{i = 1}^n \mathbb{I} \left[ \phi_{\bf m}(t_i) < x \right]
    \label{eq:ECDF}
\end{align}
of multiple realizations of both Metropolis and event-chain dynamics (without global-twist dynamics) in Figs~\ref{fig:ECDFs}(a)-(b), where 
$t_i$ is the Monte Carlo time at observation $i$ and the indicator function $\mathbb{I}(A)$ is one/zero if event $A$ does/does not occur.  Each ECDF then measures the number of simulation observations with global $U(1)$ phase value less than $x \in (-\pi, \pi]$, with symmetric simulations displaying small deviations from the target CDF $F(x) := \mathbb{P}(\phi_{\bf m} < x)$ of the uniform distribution $\mathcal{U}(-\pi, \pi)$. The low-temperature Metropolis simulations in Figs~\ref{fig:ECDFs}(a)-(b) demonstrate larger deviations from the target CDF $F(x)$ than their high-temperature counterparts, with the mean low-temperature deviations increasing with system size, and each low-temperature simulation generating a non-reproducible ECDF.  This is consistent with a loss of $U(1)$ symmetry in the low-temperature phase.  Indeed, the schematic in Fig.~\ref{fig:ECDFs}(c) presents the two possible forms of the ECDFs in the thermodynamic limit: Heaviside step functions in the low-temperature symmetry-broken phase and the target CDF $F(x)$ at high temperature.  The low-temperature Metropolis simulations in Figs~\ref{fig:ECDFs}(a)-(b) suggest that any sequence of ECDFs at any fixed simulation timescale tends to some Heaviside step function as $N \to \infty$.  Their high-temperature counterparts suggest, by contrast, symmetric convergence on the target CDF $F(x)$ on some non-divergent directional mixing timescale.  This also holds for the low- and high-temperature event-chain simulations in the inset of Fig.~\ref{fig:ECDFs}(a).  

The Cram\'{e}r-von Mises mean square distance~\cite{Cramer1928OnTheComposition,vonMises1928Wahrscheinlichkeit} 
\begin{align}
    \omega_{\phi_{\bf m},n}^2 := \int \left[ F_{\phi_{\bf m},n}(x) - F(x) \right]^2 dF(x) 
    \label{eq:CvM}
\end{align}
between $F_{\phi_{\bf m},n}(x)$ and the target CDF $F(x)$ measures the deviations from the target CDF $F(x)$, providing access to the (directional mixing) timescale on which symmetry is achieved via local dynamics.  Fig.~\ref{fig:SymmetryBreaking}(a) shows estimates of the Cram\'{e}r-von Mises statistic $\langle n \omega_{\phi_{\bf m},n}^2 \rangle $ as a function of temperature and system size.  We use supplemental global-twist dynamics to improve the statistics at high temperature~\footnote{The timescale on which symmetry is achieved via global dynamics (at arbitrarily large system size) is very long (and at most weakly $N$-dependent) compared to that achieved via local dynamics for the simulations presented here.}.  For large enough $n$ and within the simulation error, the data converge on the displayed values and indicate that $\langle \tau \omega_{\phi_{\bf m},n}^2 \rangle  \sim N$ for low-temperature Metropolis dynamics and that $\langle \tau \omega_{\phi_{\bf m},n}^2 \rangle $ is system-size independent for both event-chain dynamics (sufficiently far from the transition) and high-temperature Metropolis dynamics. 
Since $\langle \omega_{\phi_{\bf m},n}^2 \rangle \to 0$ as $\tau \to \infty$ and $\langle \tau \omega_{\phi_{\bf m},n}^2 \rangle$ converges on the directional mixing timescale $\tau_{\rm mix}$, it follows that $\langle \omega_{\phi_{\bf m},n}^2 \rangle \propto \tau_{\rm mix} / \tau$ for all $\tau > \tau_{\rm mix}$.  The data therefore imply that $\tau_{\rm mix} \sim N^{z / 2}$, where at low temperature $z = 2$ for the physical Metropolis dynamics and $z = 0$ for the symmetric event-chain dynamics (reminiscent of previous investigations~\cite{Lei2018IrreversibleMarkovChains}).  The data also imply that $z = 0$ at high temperature in both cases.  All data sets meet at a high-temperature plateau, reflecting the flattening of the Boltzmann distribution with increasing temperature.  
The event-chain results decrease with temperature, reflecting improved directional mixing (see Appendix~\ref{app:Algorithms}).

Fig.~\ref{fig:SymmetryBreaking}(a)~(inset) indicates that the Metropolis data sets of successive system sizes intersect near the transition.  This is due to the finite-size transition temperature consisting of its thermodynamic value and an additive term $\propto \!  1 / \! \left( \ln{N} \right)^2$~\cite{Bramwell1993Magnetization,Chung1999EssentialFiniteSizeEffect}.  Defining the intercept temperature $1 / [\beta_{\rm int}(N) J]$ to be the lowest temperature at which $\langle \omega_{\phi_{\bf m},n}^2(N) \rangle = \langle \omega_{\phi_{\bf m},n}^2(N / 4) \rangle$, we therefore perform local fittings of $\langle n \omega_{\phi_{\bf m},n}^2 \rangle$, with the resultant intercepts marked by grey lines in Fig.~\ref{fig:SymmetryBreaking}(a)~(inset).  The intercept temperatures are then plotted against $1 / \! \left( \ln{N} \right)^2$ in Fig.~\ref{fig:SymmetryBreaking}(b).  Fitting a straight line to the data, the reduced intercept temperature $\tilde{\beta}_{\rm BKT} / \beta_{\rm int}(N)$ extrapolates to $0.998 \pm 0.007$ in the thermodynamic limit, i.e., $\beta_{\rm int}(N) \to \beta_{\rm BKT}$ as $N \to \infty$, with $\tilde{\beta}_{\rm BKT} := 1 / (0.887 J) \simeq \beta_{\rm BKT}$.  Moreover, the estimated intercept values scale approximately with $N$ [see Fig.~\ref{fig:SymmetryBreaking}(b)~(inset)].  In conjunction with the low-temperature Metropolis results in Fig.~\ref{fig:SymmetryBreaking}(a), this demonstrates that $\tau_{\rm mix} \sim N$ for all $\beta > \beta_{\rm BKT}$.  Recalling the assumption that $\langle s_{\phi_{\bf m}}^2 \rangle \propto \tau / \tau_{\rm mix}$ for large enough $\tau < \tau_{\rm mix}$, it follows that $\langle s_{\phi_{\bf m}}^2 \rangle \sim N^{-1}$ for all $\beta > \beta_{\rm BKT}$.  This confirms Eq.~\eqref{eq:SingularLimit} and the singular thermodynamic limit of the phase fluctuations ($\langle s_{\phi_{\bf m}}^2 \rangle^{1 / 2}$) at all $\beta > \beta_{\rm BKT}$, due to nonzero long-time directional stability.  
In addition, the phase fluctuations are asymptotically smaller than the expected norm throughout the low-temperature phase, where $1 / \mathbb{E}\| {\bf m} \|$ is $O \left(N^{1 / 16}\right)$~\cite{Bramwell1993Magnetization}.  

Topological nonergodicity therefore induces broken $U(1)$ symmetry in the low-temperature BKT phase.  Symmetry is broken because the algebraic correlations combine with the diffusive Brownian spin dynamics to provoke a divergence (with system size) of the (directional mixing) timescale on which the directional phase of the $U(1)$ order parameter ergodically explores $(-\pi, \pi]$, but symmetry can be restored by non-physical global-twist dynamics that tunnel through the $U(1)$ sombrero potential and also restore topological ergodicity~\cite{Faulkner2015TSFandErgodicityBreaking,Faulkner2017AnElectricFieldRepresentation}.  The scaling of the directional mixing timescale implies that the low-temperature $U(1)$ phase fluctuations go to zero in the thermodynamic limit while being asymptotically smaller than the expected norm of the order parameter.  This case is distinct from spontaneous symmetry breaking as it cannot be identified via a singular limit of the expected $U(1)$ order parameter, leading us to define {\it general symmetry breaking} as directional (phase) fluctuations going to zero in the thermodynamic limit while being asymptotically smaller than the expected norm.  This includes spontaneous symmetry breaking and corresponds to the order parameter arbitrarily choosing some well-defined direction in the thermodynamic limit, reflecting all cases of strictly decreasing directional (phase) fluctuations being negligible compared to the expected norm in arbitrarily large systems.  

In contrast with the above Metropolis conclusions, event-chain simulations are $U(1)$-symmetric on non-divergent timescales at all nonzero temperatures.

\section{Discussion}
\label{sec:Discussion}

Previous static BKT studies focused on the expected norm of the $U(1)$ order parameter in both thermodynamic~\cite{Mermin1966AbsenceFerromagnetism,Hohenberg1967ExistenceOfLong-RangeOrder} and finite~\cite{Archambault1997MagneticFluctuations,Bramwell1998Universality} systems, where the latter led to much success in modelling experimental quantities that are functions of the expected norm~\cite{Bishop1978StudySuperfluid,Bramwell2015PhaseOrder,Bramwell1993Magnetization,Huang1994MagnetismFewMonolayersLimit,Elmers1996CriticalPhenomenaTwoDimensionalMagnet,Chung1999EssentialFiniteSizeEffect,BedoyaPinto2021Intrinsic2DXYFerromagnetism,Venus2022RenormalizationGroup}.  Our dynamical study, by contrast, explicitly demonstrates that the low-temperature order parameter arbitrarily chooses some well-defined \emph{direction} in the thermodynamic limit.  This asymptotically slow directional mixing predicts negligible $U(1)$ phase fluctuations compared to the strictly decreasing expected norm in arbitrarily large experimental BKT systems, thus providing a theoretical framework for broken symmetry to the thermodynamic limit throughout the low-temperature BKT phase.  This constitutes a model for the directional mixing (or memory) timescale $\tau_{\rm mix} \sim N$.  We suggest experiments on a single Josephson junction formed from two nodes of 2D superconducting film as an ideal showcase of the phenomenon.  Moreover, measurements of the magnetization vector in XY magnetic films (with a six-fold crystal field~\cite{Jose1977Renormalization}) should provide further direct experimental evidence, as should the orientational order parameter in the hexatic phase of colloidal films~\cite{Bernard2011TwoStepMelting,Thorneywork2017TwoDimensionalMelting}.  Indeed, this paper provides the full dynamical framework for system-spanning symmetry-broken spin/condensate-phase coherence in the wide array of critical BKT experimental systems~\cite{Baity2016EffectiveTwoDimensionalThickness,Shi2016EvidenceCorrelatedDynamics,Bishop1978StudySuperfluid,Bramwell2015PhaseOrder,Wolf1981TwoDimensionalPhaseTransition,Resnick1981KosterlitzThouless,Bramwell1993Magnetization,Huang1994MagnetismFewMonolayersLimit,Elmers1996CriticalPhenomenaTwoDimensionalMagnet,BedoyaPinto2021Intrinsic2DXYFerromagnetism,Hadzibabic2006KosterlitzThouless,Fletcher2015ConnectingBKTAndBEC,Christodoulou2021Observation,Thorneywork2017TwoDimensionalMelting}, motivating experiments on BKT systems in general.  

Our results also imply that the strongly nonergodic electrical-resistance PDFs measured in the LSCO films~\cite{Shi2016EvidenceCorrelatedDynamics} are likely to be due to increasingly large regions of symmetry-broken condensate-phase coherence as the BKT transition is approached from high temperature, described by a critical slowing down of the symmetry-breaking order parameter analogous to that of the 2D Ising model.  This is because increasingly large symmetry-broken regions persisting on significant nonergodic timescales are a necessary precursor to the system-spanning symmetry-broken region that we have demonstrated at low temperature.  
We leave the remainder of this hypothesis to a separate publication and suggest that this second effect should also be detectable near the BKT transition in 2D Josephson-junction arrays and XY magnetic films, and similarly near the solid--hexatic transition in colloidal films.  Indeed, our hypothesis also applies to 3D superconductors, Josephson-junction arrays and XY magnets, but the 2D case is particularly significant to experiment as the BKT transition occurs over a broad temperature range in finite systems~\cite{Bramwell1993Magnetization,Chung1999EssentialFiniteSizeEffect}.  

We additionally showed that event-chain 2DXY simulations are $U(1)$-symmetric on non-divergent timescales at all nonzero temperatures.  This is analogous to the accelerated mixing (relative to Metropolis) of Swendsen--Wang/Wolff simulations~\cite{Swendsen1987Nonuniversal,Wolff1989Collective} of the 2D Ising model at low temperature, but due in this case to the deterministic event-chain dynamics efficiently exploring the order-parameter distribution along low-gradient directions, with our event-chain simulations also suggesting significantly improved exploration of the heavy tail of the asymmetric $\| {\bf m} \|$ distribution that develops at low temperature~\cite{Archambault1997MagneticFluctuations}.  This fundamental characteristic of event-chain Monte Carlo suggests that it may also alleviate critical slowing down by advancing deterministically through the flattened regions of the order-parameter distribution typically found at continuous phase transitions -- possibly related to the superdiffusive dynamics of the location of the 2DXY active spin~\cite{Kimura2017AnomalousDiffusion}, and again in analogy with the Swendsen--Wang/Wolff simulations.  This hypothesis is also suggested by the $\sqrt{N}$ speed-up of lifted (relative to standard) Metropolis--Hastings simulations of the Curie--Weiss model at its phase transition~\cite{Bierkens2017APiecewiseDeterministic}, as its limiting behavior~\cite{Bierkens2017APiecewiseDeterministic} is a similar piecewise deterministic Markov process that originated in Bayesian computation.  On the contrary, it may be that event-chain Monte Carlo (as presented here) cannot overcome the impact of the vortex excitations~\cite{Lei2018IrreversibleMarkovChains}.  
If so, it may then prove to be a powerful tool for separating the vortex and spin-wave contributions, and it would be interesting to explore whether \emph{factor fields}~\cite{Lei2019ECMCwithFactorFields} can circumvent those from the vortices (as factor fields typically lead to accelerated event-chain mixing~\cite{Lei2019ECMCwithFactorFields}).
This motivates additional studies into the power of piecewise deterministic Markov processes at phase transitions across statistical physics, Bayesian computation and applied probability.  Indeed, the foundational viewpoint of broken symmetry presented here -- an asymptotically slow mixing between equilibrium regions of equal (measure and) probability mass -- could be viewed as the canonical choice in Bayesian computation and applied probability.  This cross-pollination of knowledge may motivate further innovations across all three fields.

\begin{figure*}[t]
  \hspace{-0.6em}
  \includegraphics[width=\linewidth]{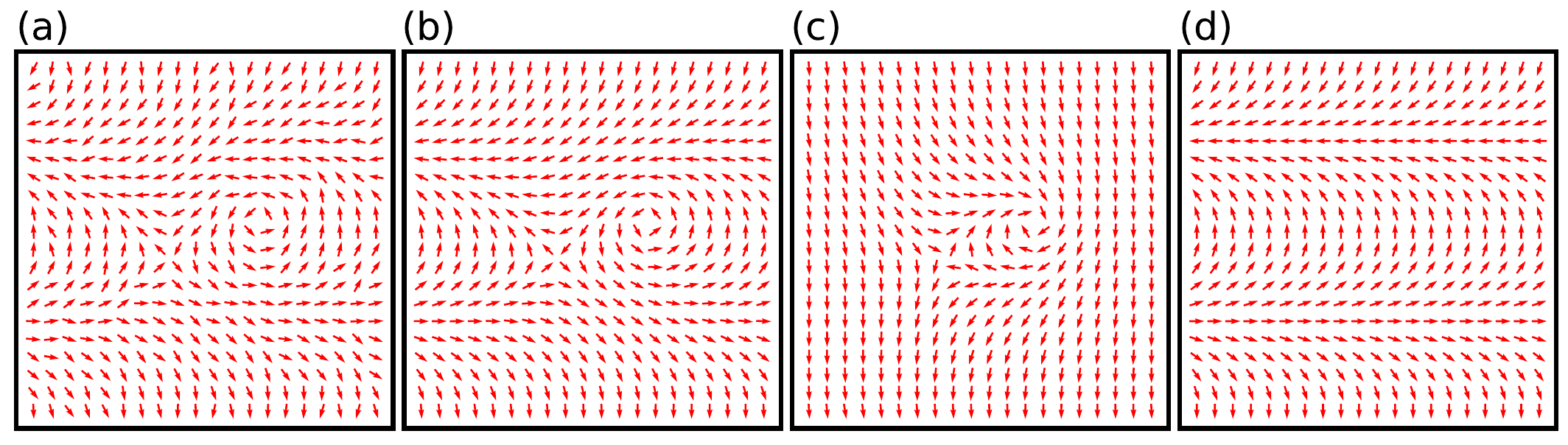}
  \caption{Each spin configuration is composed of local topological defects (spin vortices), global topological defects (internal global spin twists) and continuous fluctuations around these topological defects (spin waves).  Red arrows represent spins.  (a) Typical 2DHXY configuration at $\beta J = 1.5$ with fixed topological defects [i.e., for illustrative purposes, we fixed two local defects (and one global defect) in position and allowed spin waves to fluctuate around this constrained configuration -- this is distinct from the non-constrained simulations presented in the main body of the paper].  (b) Zero-temperature minimization of (a), i.e., with spin waves annealed away.  (c)-(d) Configuration in (b) split into its (c) vortex and (d) internal global-twist components.}
  \label{fig:TopDefects}
\end{figure*}

We also introduced \emph{general symmetry breaking}.  While both fall within this general concept, the present result is distinct from spontaneous symmetry breaking because it cannot be identified by taking the thermodynamic and then zero-symmetry-breaking-field limits of the expected order parameter, 
though this elegant mathematical formalism~\cite{Mermin1966AbsenceFerromagnetism,Hohenberg1967ExistenceOfLong-RangeOrder} of singular limits (of statistical expectations) paved the way to the subtleties of the thermodynamic limit in a critical system~\cite{Archambault1997MagneticFluctuations,Bramwell1998Universality}.  Indeed, the vanguard of these first and second waves of BKT theory correctly identified both an absence of spontaneous symmetry breaking and an experimentally relevant order-parameter norm, both of which are united by our general concept as it allows the expected norm to go to zero in the thermodynamic limit provided the phase fluctuations are asymptotically smaller.  We note that, as $\mathbb{E}\|\widetilde{{\bf m}}\|$ is system-size independent, it is tempting to define the broader concept with respect to $\widetilde{{\bf m}}$ and without stating ``while being asymptotically smaller than the expected norm''.  However, cases of phase fluctuations being asymptotically greater than or equal to $\mathbb{E}\| {\bf m} \|$ must be excluded from general symmetry breaking, which might not be satisfied by this adaptation.  We add that the present work suggests a symmetry-breaking singular limit of $\mathbb{E}\widetilde{{\bf m}}$ at low temperature, but preliminary simulation results suggest that $\mathbb{E}\|\widetilde{{\bf m}}\|$ is nonzero for all finite $\beta$ in zero field.  It would be interesting to explore whether $1 / \lim_{N \to \infty} \mathbb{E} \|\widetilde{{\bf m}}({\bf h} = {\bf h}_0)\|$ (or potentially $\lim_{N \to \infty} \sigma_{\|{\bf m}({\bf h} = {\bf h}_0)\|} / \lim_{N \to \infty} \mathbb{E} \|{\bf m}({\bf h} = {\bf h}_0)\|$) can be made arbitrarily  small with decreasing $\| {\bf h}_0 \|$ at low temperature.

\acknowledgements
It is a pleasure to thank S.~T. Bramwell, P.~C.~W. Holdsworth, Z. Shi, D. Popovi\'{c}, S.~Livingstone and A.~C. Maggs for many illuminating discussions.  The author is grateful to STB and ZS for detailed comments on the manuscript, to PCWH, V.~Kaiser, J.~F.~Annett and M.~V. Berry for comments on the manuscript, to JFA for discussions on superfluid films, to MVB for the analogy with quantum chaos, and to W. Krauth for sharing his 2DXY event-chain code.  The author acknowledges support from EPSRC fellowship EP/P033830/1.  Simulations were performed on BlueCrystal 4 at the Advanced Computing Research Centre (University of Bristol) who granted us an exceptional 30TB of scratch storage for Fig.~\ref{fig:SymmetryBreaking}.

\appendix

\section{Code and data availability}

Code is freely available at \href{https://github.com/michaelfaulkner/xy-type-models}{https://github.com/michaelfaulkner/xy-type-models}, commit hash \href{https://github.com/michaelfaulkner/xy-type-models/commit/acd5334ee30923f03d8d95d53ce5ea84302b618e}{acd5334}.  All published data can be reproduced using this application (as outlined in its README) and are available at the University of Bristol data repository, \href{https://data.bris.ac.uk/data/}{data.bris}, at \href{https://doi.org/10.5523/bris.3ov1rl6xtshwv2iuixrbs6f39q}{https://doi.org/10.5523/bris.3ov1rl6xtshwv2iuixrbs6f39q}.

\section{Field decomposition and topological ergodicity}
\label{app:TopologicalErgodicity}

Supplemental global-twist dynamics were used in Figs~\ref{fig:GlobalTwists} and \ref{fig:SymmetryBreaking}.  We elucidate the global twists via the 2D harmonic XY (2DHXY) model~\cite{Vallat1992ClassicalFrustratedXY,Bramwell1994Magnetization} -- a piecewise-parabolic analogue of the 2DXY model whose quadratic zero-field potential $J \sum_{\langle i,j \rangle} \left( \Delta \theta_{ij} \right)^2 / 2$ maintains the $2\pi$ XY periodicity via the definition $\Delta \theta_{ij} := \left( \varphi_i - \varphi_j + \pi \right) \!\! \mod (2\pi) - \pi$ while mapping directly to the 2D lattice-field electrolyte~\cite{Vallat1994CoulombGas,Faulkner2017AnElectricFieldRepresentation}.  Each spin configuration decomposes into three excitations: local topological defects (spin vortices), global topological defects (internal global spin twists) and continuous 
fluctuations around these defects (spin waves).  An example 2DHXY configuration at $\beta J = 1.5$ and with a fixed topological-defect configuration is shown in Fig.~\ref{fig:TopDefects}(a) [i.e., for illustrative purposes, we fixed two local defects (and one global defect) in position and allowed spin waves to fluctuate around this constrained configuration -- this is distinct from the non-constrained simulations presented in the main body of the paper].  Fig.~\ref{fig:TopDefects}(b) is the zero-temperature 
minimization of this configuration: topological defects are fixed and spin waves are annealed away.  Fig.~\ref{fig:TopDefects}(c) depicts Fig.~\ref{fig:TopDefects}(b) with global spin twists [defined in Eq.~\eqref{eq:ExternalGlobalTwists}] applied along each Cartesian dimension until the potential is minimized by some ${\bf q} \in \mathbb{Z}^2$, leaving behind the vortex field.  The right/left-hand local topological defect is a positive/negative vortex, about which the spins rotate by $\pm 2\pi$.  ${\bf q}$ is identified with the \emph{global twist-relaxation field} $\tilde{{\bf t}}$, which removed the internal global spin twist (described by the \emph{internal global twist field} ${\bf t} = - \tilde{{\bf t}}$) depicted in Fig.~\ref{fig:TopDefects}(d).  Internal global spin twists are global topological defects.  Each can be generated by a vortex tracing a closed path around the torus (in the $x$ direction in this case) before annihilating another of opposite sign~\cite{Faulkner2015TSFandErgodicityBreaking,Faulkner2017AnElectricFieldRepresentation} (see, in particular, Fig.~5 of Ref.~\cite{Faulkner2017AnElectricFieldRepresentation}).  In the emergent electrostatic-field representation in which the positive/negative vortex maps to a positive/negative emergent charge, the vortex, spin-wave and internal global-twist components map to~\cite{Faulkner2017AnElectricFieldRepresentation} (respectively) the low-energy solution to the Gauss law for the emergent charges, the purely rotational auxiliary gauge field of the 2D lattice electrolyte~\cite{Maggs2002LocalSimulationAlgorithms,Faulkner2015TSFandErgodicityBreaking} and the topological sector ${\bf w} \in \mathbb{Z}^2$ of the 2D electrolyte, with ${\bf w} = (t_y, -t_x)$~\cite{Faulkner2015TSFandErgodicityBreaking}.  The same decomposition recipe defines the global twist-relaxation field $\tilde{{\bf t}}$ in the 2DXY model, but this mapping to the electrolyte-field components is then only approximate.  One may circumvent this by using the 2DHXY potential in the decomposition recipe.  This defines the local and global topological defects, but the global topological defects will not always correspond to the 
global twist-relaxation field.  

In Figs~\ref{fig:GlobalTwists} and \ref{fig:SymmetryBreaking}, we supplemented local 2DXY dynamics (described in Appendix~\ref{app:Algorithms}) with externally applied (to the non-decomposed spin field) global spin twists.  Due to the coupling between each component of even the 2DHXY spin field, the global twist-relaxation field $\tilde{{\bf t}}$ cannot be isolated and uniquely manipulated by these global-twist dynamics, as such dynamics may alter all three spin excitations.  In contrast, topological-sector dynamics alter only the topological sector of the 2D electrolyte, because the auxiliary gauge field is independent of the remaining electric field~\cite{Maggs2002LocalSimulationAlgorithms,Faulkner2015TSFandErgodicityBreaking}.  However, Fig.~\ref{fig:GlobalTwists}(b) shows that, as for the case of topological-sector events in the 2D electrolyte~\cite{Faulkner2015TSFandErgodicityBreaking}, the probability of global-twist events is system-size independent sufficiently far from the transition, and non-negligible at all system sizes and nonzero temperatures, despite being small at low temperature.  Defining \emph{topological order}/\emph{nonergodicity} under some dynamics by vanishing $\tilde{{\bf t}}$ fluctuations $\left( \langle s_{\tilde{{\bf t}}}^2 \rangle^{1 / 2} \right)$ in the thermodynamic limit, \emph{topological ergodicity} then corresponds to 
\begin{align}
    \lim_{\tau \to \infty} \lim_{N \to \infty} \langle s_{\tilde{{\bf t}}}^2 (\beta, \tau, N) \rangle = {\rm Var}[\tilde{{\bf t}}] 
\end{align}
and global-twist dynamics ensure topologically ergodic simulations on non-divergent timescales (with topologically ergodic simulations defined by a simulation variance $s_{\tilde{{\bf t}}}^2$ that has converged to its expected value ${\rm Var}[\tilde{{\bf t}}]$ within some small $\varepsilon > 0$, reflecting ergodic exploration of the global twist-relaxation field $\tilde{{\bf t}} \in \mathbb{Z}^2$).  This is analogous to topological-sector dynamics ensuring topologically ergodic simulations (on non-divergent timescales) with respect to $\mathbf{w}$ in the 2D electrolyte~\cite{Faulkner2015TSFandErgodicityBreaking}.  Since, to leading order, topologically ergodic simulations require ergodic exploration of $\tilde{{\bf t}}$ only over the five-element set $\{ (0, 0), \pm (1, 0), \pm (0, 1) \} \subset \mathbb{Z}^2$, it is reasonable to assume that these timescales are 
on the order of the non-divergent timescales on which symmetric Metropolis simulations are ensured by supplemental global-twist dynamics ($r \in \mathbb{N}$ of the tunnelling events described in Section~\ref{sec:GlobalTwistDynamics} result in $r$ of the Dirac distributions also described there).  
Global-twist dynamics also ensure ergodic simulations on non-divergent timescales with respect to a non-annealed analogue of the global twist-relaxation field.  This will be set out in detail in a future review article~\cite{Faulkner2024BKTSymmetryReview}.

\section{Simulation and fitting algorithms}
\label{app:Algorithms}

\subsection{Local Metropolis Monte Carlo}

We used the local Metropolis algorithm with Gaussian noise to investigate the 2DXY model with Brownian spin dynamics.  Each Metropolis proposal attempts to perturb a single spin by an amount $\varepsilon \sim \mathcal{N}(0, \sigma_{\rm noise})$, with $\sigma_{\rm noise}^2$ the variance of the Gaussian-noise distribution.  Such proposals are accepted with probability $\min \left[ 1, \exp (-\beta \Delta U) \right]$, with $\Delta U$ the potential of the proposed configuration minus its current value.  For a set of linearly coupled 1D harmonic oscillators, Metropolis dynamics were proven rigorously~\cite{Neal2006OptimalScalingForPartially} to converge on Brownian dynamics (in the thermodynamic limit) with emergent physical time step $\Delta t_{\rm phys} := a_{\rm Metrop} \sigma_{\rm noise}^2 \Delta t_{\rm Metrop} / 2$ per unit (Brownian) diffusivity.  Here, $a_{\rm Metrop}$ is the Metropolis acceptance rate and $\Delta t_{\rm Metrop}$ is the Metropolis Monte Carlo time step, which we define as the elapsed Monte Carlo time between $N$ attempted single-spin moves.  We tune $\sigma_{\rm noise}$ such that $a_{\rm Metrop} \simeq 0.6$ throughout and draw observations after each Metropolis Monte Carlo time step.  

\subsection{Event-chain Monte Carlo}

We benchmarked the diffusive local Metropolis dynamics against the event-chain Monte Carlo algorithm~\cite{Michel2015EventChain}.  Event-chain Monte Carlo inverts Metropolis: rather than proposing a single discrete spin translation before drawing a random number to decide whether to accept the move, event-chain Monte Carlo draws the random number and then continuously advances some \emph{active spin} at fixed (anticlockwise) velocity $v > 0$ until a Metropolis rejection would have occurred at some \emph{particle-event time} $t_\eta > 0$.  

To sample the particle-event times, we consider a sequence of $m$ proposed Metropolis translations of length $\delta > 0$ in the positive (i.e., anticlockwise) direction, starting from some initial time $t_0 \ge 0$.  Defining $\Delta U_{t_0,i} := U \left[ \varphi_1(t_0), \dots , \varphi_a(t_0) + i \delta, \dots , \varphi_N(t_0) \right] - U \left[ \varphi_1(t_0), \dots , \varphi_a(t_0) + (i - 1) \delta, \dots , \varphi_N(t_0) \right]$ and assuming $\varphi_a(t_0) + m \delta < \pi$ (as $\varphi_k \in ( -\pi, \pi ]$ for all spins $k$), the probability of accepting all proposals and translating the active spin $a$ through the total distance $\eta := m \delta$ is 
\begin{align}
    p_{t_0, \eta} = & \prod_{i = 1}^{\eta / \delta} \min \left[ 1, \exp (-\beta \Delta U_{t_0, i}) \right] \nonumber\\
    = & \exp \left[ -\beta \sum_{i = 1}^{\eta / \delta} \max \left( 0, \Delta U_{t_0, i} \right) \right] . \nonumber
\end{align}
Defining $\boldsymbol{\varphi} := (\varphi_1, \dots , \varphi_N)$, it follows that 
\begin{align}
    p_{t_0, \eta} \to \exp \left[ -\beta \int_{\varphi_a(t_0)}^{\varphi_a(t_0) + \eta} \max \left[ 0, \partial_a U(\boldsymbol{\varphi}') \right] d\varphi_a' \right] \nonumber
\end{align}
as $\delta \to 0$ with $\eta$ fixed, i.e., in the continuous-time limit.  We therefore draw some random number $\Upsilon \sim \mathcal{U}(0, 1]$ at $t_0$ and if 
\begin{align}
    -\ln \Upsilon < \beta \int_{\varphi_a(t_0)}^\pi \max \left[ 0, \partial_a U(\boldsymbol{\varphi}') \right] d\varphi_a' ,
    \label{eq:BoundaryEventECMC}
\end{align}
we solve 
\begin{align}
    -\ln \Upsilon = \beta \int_{t_0}^{t_\eta} \max \left[ 0, v \, \partial_a U(\boldsymbol{\varphi}(t)) \right] dt 
    \label{eq:MasterEquationECMC}
\end{align}
to sample the next particle-event time $t_\eta = t_0 + \eta / v$.  This defines a \emph{particle event} and one of the four (nearest) neighboring spins $k$ then becomes active with probability $\propto \max \left[ 0, - v \, \partial_k U(\boldsymbol{\varphi}(t = t_\eta)) \right]$.  If Eq.~\eqref{eq:BoundaryEventECMC} does not hold, however, then $\varphi_a(t = t_{\rm b}) = \pi$ at the \emph{boundary-event time} $t_{\rm b} := t_0 + [\pi - \varphi_a(t_0)] / v$.  In such instances, the active spin is instantaneously translated to $-\pi$, defining a \emph{boundary event} at $t_{\rm b}$.  The active spin remains active following this form of \emph{teleportation portal}~\cite{Bierkens2023PDMPsWithBoundaries}.  

Event-chain Monte Carlo therefore amounts to sampling from a non-homogeneous Poisson process with intensity function (or \emph{event rate}) $\beta \max \left[ 0, v \, \partial_a U(\boldsymbol{\varphi}(t)) \right]$.  In practice, however, it is typically challenging to solve Eq.~\eqref{eq:MasterEquationECMC}.  In our simulations, we instead sample one particle/boundary-event time per two-particle potential and define the next event as that corresponding to the soonest of these.  If this is a particle event, the corresponding vetoing spin becomes active.  Observations are then drawn every $N$ units of event-chain time, though one may alternatively choose the observation times via some homogeneous Poisson process with intensity function proportional to $N$~\footnote{It was erroneously stated in the published version~\cite{Faulkner2024SymmetryBreakingBKT} that observations were drawn at every $N^{\rm th}$ particle event (which would have resulted in a biased sample of $\| {\bf m} \|$ but not of $\phi_{\bf m}$).}.  In lower dimensional models, practitioners may instead store the particle event times $\{ t_\eta^j \}$ and corresponding system states $\{ \boldsymbol{\varphi}(t_\eta^j) \}$ in order to reconstruct the entire Markov process.

Finally, in the description of Fig.~\ref{fig:SymmetryBreaking}(a) in Section~\ref{sec:SymmetryBreaking}, we stated that the event-chain Cram\'{e}r-von Mises statistic $\langle n \omega_{\phi_{\bf m},n}^2 \rangle$ decreasing with temperature reflects improved directional mixing at lower temperatures.  We hypothesize that this is due to increased long-range spin--spin correlations with decreasing temperature.  

\subsection{Other simulation details}

Simulations for Figs~\ref{fig:MagnetisationEvolution}, \ref{fig:GlobalTwists}(a) and \ref{fig:ECDFs} started from randomized initial configurations.   Those for Figs~\ref{fig:GlobalTwists}(b)-(c) and \ref{fig:SymmetryBreaking} started from ordered initial configurations.  $10^4$ and $10^5$ initial equilibration observations were discarded (respectively) in each event-chain and Metropolis simulation.  Standard error estimates were used (i.e., not accounting for autocorrelation in the Markov chain) and all non-visible error bars are smaller than the marker size.  We consider simulations to have directionally mixed (i.e., $\langle n \omega_{\phi_{\bf m},n}^2 \rangle$ to have converged) if $\langle s_{\phi_{\bf m}}^2 \rangle > 0.98 {\rm Var}[\phi_{\bf m}]$.

\subsection{Polynomial-fitting algorithms}

For the local fittings in Fig.~\ref{fig:SymmetryBreaking}(a)~(inset), we applied natural logarithms to each data set (corresponding to each system size) and then performed second-order polynomial fittings (within each data set) to the three data points nearest to each intercept temperature (defined in Section~\ref{sec:SymmetryBreaking}).  Estimated Monte Carlo errors were used in the fittings.  We then performed first-order fittings to a) the resultant intercept temperatures vs $1 / \left( \ln N \right)^2$, and b) the natural logarithm of the resultant intercept values vs $\ln N$.  We used the NumPy~\cite{Harris2020Numpy} package polyfit for all fittings. 

\bibliographystyle{ieeetr}
\bibliography{Faulkner}

\end{document}